\documentclass[aps,12pt,superscriptaddress,preprintnumbers,
                secnumarabic]{revtex4}

\usepackage{epsfig}

\begin{document} 

\title{\Large Perturbative and Nonperturbative Contributions to a Simple
Model for Baryogenesis}

\author{K.~R.~S.~Balaji}
\email[]{balaji@hep.physics.mcgill.ca}
\affiliation{
Dept.~of Physics, McGill University, 3600 University Street, Montr´eal QC, H3A 2T8, 
Canada}
\affiliation{
Physique des Particules, Universit\'e
de Montr\'eal,C.P. 6128, succ. centre-ville, Montr\'eal, QC,
Canada H3C 3J7}
\affiliation{D\'epartement des Sciences de la Terre et de
l'Atmosph\`ere, Universit\'e du Qu\'ebec \`a Montr\'eal,
C.P. 8888, succ. centre-ville, Montr\'eal, QC, Canada H3C 3P8}

\author{Bj\"orn~Garbrecht}
\email[]{bjorn@hep.man.ac.uk}
\affiliation{School of Physics \& Astronomy, University of Manchester, Oxford Road, Manchester M13 9PL, UK}



\preprint{MCGILL-01-06}
\preprint{MAN/HEP/2006/4}

\begin{abstract}
Single field baryogenesis, a scenario for Dirac leptogenesis sourced by
a time-dependent scalar condensate, is studied on a toy model.
We compare the creation of
the charge asymmetry by the perturbative decay
of the condensate with the nonperturbative decay, a process of particle
production commonly known in the context of inflation as preheating.
Neglecting backreaction effects, we find that over a wide parametric range
perturbative decay and
preheating contribute by the same order of magnitude to the baryon asymmetry.
\end{abstract}


\maketitle

\section{Introduction}

Models for baryogenesis tie together cosmology and particle
physics~\cite{baryo}. The discovery of small neutrino masses~\cite{sk}
and their explanation \emph{via} lepton-number violating
Majorana masses and the see-saw mechanism strongly supports the
leptogenesis mechanism~\cite{lepto}.
An alternative to this scenario is to assume pure
Dirac mass terms for the neutrinos, that is the absence of Majorana masses,
and to induce an asymmetry between the left- and right-handed neutrinos, 
which is subsequently turned into baryon number through sphaleron transitions.
This idea is referred to as
Dirac leptogenesis~\cite{DickLindnerRatzWright:1999}.

In many scenarios for baryogenesis the necessary $C$ and $CP$ violation
occur simultaneously, induced by a matrix of Yukawa couplings.
An exception is electroweak baryogenesis~\cite{KuzminRubakovShaposhnikov:1985,KonstandinProkopecSchmidt:2005},
where in first place a $CP$-violating axial asymmetry is produced which is
subsequently not erased {\it via} equilibration. Parity violation
is then contributed in a second step due to sphaleron interactions.

Also Dirac leptogenesis relies on the transition of an
axial asymmetry into baryons through the sphaleron process.
The initial asymmetry is stored within Dirac neutrinos and does not get
erased due to equilibration until electroweak symmetry breaking
since the Yukawa coupling to the Standard Model Higss field is tiny.
While the $CP$ asymmetry can be provided through a matrix of Yukawa couplings
and the out-of equilibrium decay of heavy scalar particles
into Dirac neutrinos~\cite{DickLindnerRatzWright:1999},
it has been suggested that also a single Dirac mass term can
source $CP$, provided it is time dependent.
This mass term can arise due to Yukawa couplings of neutrinos to
a rolling scalar field, and the resulting mechanism has been named
single field baryogenesis~\cite{BalajiBrandenberger:2005}.
Recently, an interesting realisation of this mechanism through a decaying
Affleck Dine condensate has been proposed~\cite{AbelPage:2006}.

Interpreting the scalar condensate oscillating around \emph{zero} as
a large amount of scalar quanta at zero momentum, the axial asymmetry
can be generated due to the perturbative decay of these particles,
as commonly assumed in scenarios for
Affleck-Dine baryogenesis~\cite{AffleckDine:1984}.
However particles can also be produced nonperturbatively, as first pointed
out in Ref.~\cite{TraschenBrandenberger:1990}, a process which is
often referred to as preheating in the context of the decay of the
inflaton~\cite{preheating}.
Various aspects of preheating from the decay of flat directions are discussed
in~\cite{PostmaMazumdar:2003}.

In parallel, in the coherent baryogenesis~\cite{GarbrechtProkopecSchmidt:2004}
scenario, the oscillating condensate
leads directly and at tree-level to the production of a charge asymmetry
during preheating
when it couples to matter such that a time-dependent $C$ and $CP$ violating
mass matrix arises. Consequently, in the case of a single time-dependent
mass term a preheating process can lead to an
axial asymmetry~\cite{GarbrechtProkopecSchmidt:2002} and thereby source single field
baryogenesis.

In the present analysis, we focus on the importance of non-perturbative
contributions to the baryon asymmetry, and we choose the single field model
due to its simplicity. We emphasise nonetheless that
nonperturbative particle production may be of relevance for other
scenarios, {\it e.g} Affleck-Dine baryogenesis.

\section{Perturbatively sourced single field baryogenesis}
Let us begin by considering a simple toy-model potential 
\begin{eqnarray}
\label{Potential:Interactions}
V=
\frac{\mu^2}{2}\left(|\phi_u|^2+|\phi_d|^2\right)
+\frac{m^2}{2}\left(\phi_u \phi_d +\phi_u^* \phi_d^*\right)
+\lambda_\nu L \phi_u \bar\nu_R+\lambda_\nu \bar L \phi_u^* \nu_R~
.
\end{eqnarray}
The fields $\phi_u$ and $\phi_d$ are scalar and are multiplets
of the electroweak group $G_{EW}={\rm SU}(2)_L \times {\rm U}(1)_Y$,
$\phi_u=\left({\bf 2},\frac 12\right)$,
$\phi_d=\left({\bf 2},-\frac 12\right)$, while
$L=\left({\bf 2},-\frac 12 \right)$
with the components
\begin{equation}
L=\left(
\begin{array}{c}
\nu_L\\
e_L
\end{array}
\right)~,
\end{equation}
 and $\nu_R=\left({\bf 1},0\right)$
are Weyl fermions.
The scalar mass eigenstates are then
$\frac{1}{\sqrt 2}\left(\Im[\phi_u]+\Im[\phi_d]\right)$ and
$\frac{1}{\sqrt 2}\left(-\Re[\phi_u]+\Re[\phi_d]\right)$, both with
mass $\sqrt{\mu^2-m^2}$, and eigenstates,
$\frac{1}{\sqrt 2}\left(-\Im[\phi_u]+\Im[\phi_d]\right)$ and
$\frac{1}{\sqrt 2}\left(\Re[\phi_u]+\Re[\phi_d]\right)$
with mass $\sqrt{\mu^2+m^2}$. The inflationary Hubble rate is given by
$H_I$, and we assume that both of these mass eigenvalues are slightly 
below this value.
Therefore, at horizon exit the scalar fields get
amplified up to a magnitude $\sim H_I$ and a random direction in
${\rm SU}(2)_L$-space. Inflationary expansion then leaves behind a homogeneous
vacuum expectation value for the scalar fields in our patch of the Universe.
This induces large neutrino masses, such that they
initially do not thermalise. In turn, the potential for $\phi_{u,d}$ does not
get altered by thermal corrections.

Since we are interested in neutrino production, in the following, we consider
only the neutral components $\phi^0_{u,d}$. In particular, since the asymmetry is
produced from $\phi^0_u$, we take this to be the source field as in the
single field baryogenesis scenario. Furthermore, we assume $m\ll\mu$.

Coherent oscillations begin at the time when the Hubble rate has decreased
to the value $\mu$, and the solution for $\phi_u^0$ can be approximated
for small small $\mu/H$ by
\begin{eqnarray}
\Re[\phi_u^0]&=&\left[A_1^R \cos\left(\sqrt{\mu^2-m^2}t\right)
             +A_2^R \cos\left(\sqrt{\mu^2+m^2}t\right)\right]a^{-3/2}(t)~,\\
\Im[\phi_u^0]&=&\left[A_1^I \cos\left(\sqrt{\mu^2-m^2}t\right)
             +A_2^I \cos\left(\sqrt{\mu^2+m^2}t\right)\right]a^{-3/2}(t)~,
\nonumber
\end{eqnarray}
where the values of $A^{R,I}_{1,2}$ are random initial values arising from
inflation, as described above, and $a(t)$ denotes the scale factor of the
Universe, $t$ denotes comoving time. In order to keep the present
discussion simple, we assume $A_1^R=A^R$,
$A_2^R=0$, $A_2^I=A^I$ and $A_1^I=0$ in our patch of the Universe.
Under these conditions, the charge density
carried by the field $\phi_u^0$ is given by
\begin{eqnarray}
\label{charge:scalar}
Q_\phi&=&\frac{\rm i}{2}
\left({\phi_u^{0^*}}\dot\phi_u^0-\phi_u^0 {{\dot\phi_u^{0^*}}}\right)
\approx a^{-3} \mu A^R A^I \sin\left(\frac{m^2}{\mu}t\right)~,
\end{eqnarray}
where we expanded in $\mu/m$ and neglected time derivatives acting on the
scale factor. According to the interaction term with the leptons in the
potential~(\ref{Potential:Interactions}), $Q_\phi$ is transferred
to a charge asymmetry within the left handed neutrinos when $\phi_u^0$
decays. Due to conservation of total lepton number, a precisely opposite amount of the
asymmetry is stored within the right handed neutrinos. However, this asymmetry in the right-handed
sector is not transferred into baryons by sphalerons due to the left-handed nature of interactions.

We assume that the Universe is radiation dominated when coherent oscillations
commence and that this remains so until the scalar fields decay,
such that they contribute only negligibly to the entropy density $s$.
Just like $Q_\phi$, $s$ scales down as $a^{-3}$, such that we find for the
asymmetry within left-handed neutrinos at the time $\Gamma^{-1}$, when
the scalar field decays,
\begin{equation}
\frac{n(\nu_L)-n(\bar\nu_L)}{s}=
\alpha
\frac{\Gamma_\nu}{\Gamma}
\frac{A^R A^I}{\mu^{1/2}m_{Pl}^{3/2}}
 \sin\left(\frac{m^2}{\mu}\Gamma^{-1}\right)~.
\label{Asymmetry:Perturbative}
\end{equation}
Here, we have used the relations $H=1.66 g_*^{1/2} T^2/m_{Pl}$~,
$s=\frac{2\pi^2}{45}g_*T^3$ and have taken $H\approx\mu$.
The number of relativistic degrees of freedom is denoted by
$g_*$, such that $\alpha$ is a numerical constant of order \emph{one} for
realistic values of $g_*$. Furthermore, we have assumed that $\phi_u^0$
decays at a total rate $\Gamma$, whereas the decay rate into neutrinos
is given by $\Gamma_\nu=\lambda_\nu^2\mu/(8\pi)$, such that a branching factor of
$\Gamma_\nu/\Gamma$ arises.

\section{Nonperturbative source}
Following Ref.~\cite{GarbrechtProkopecSchmidt:2002}, 
we calculate the axial asymmetry induced by the nonperturbative decay
of $\phi_u^0$.
We do so by solving numerically the conformally
rescaled Dirac equation
\begin{equation}
\label{Dirac:Eqn}
\left[{\rm i}\partial\!\!\!/-m_R+{\rm i}\gamma^5m_I\right]\psi=0\,.
\end{equation}
We take
\begin{equation}
\psi=\left(\begin{array}{c}
\nu_L\\
\nu_R
\end{array}\right)~,
\end{equation}
such that by the potential~(\ref{Potential:Interactions}), we have
\begin{equation}
m_R=a\lambda_\nu\Re[\phi_u^0],\quad m_I=a\lambda_\nu\Im[\phi_u^0]~.
\label{m:Phi}
\end{equation}
Furthermore, we introduce the conformal time $\eta$, which is related
to comoving time as $dt=ad\eta$, and we take $\partial_0=\partial_\eta$.

Let us introduce the positive and negative frequency
mode functions, $u_{h}(\mathbf{k},\eta)$ and 
$v_{h}(\mathbf{k},\eta) = - i\gamma^2(u_{h}(\mathbf{k},\eta))^{*}$,
respectively.
They form a basis for the Dirac field,
\begin{eqnarray}
\psi(x) \!=\! \int \frac{d^3k}{(2\pi)^3}\sum\limits_h
          {\rm e}^{\!-{\rm i}\mathbf{k}\cdot\mathbf{x}}
    \left(u_{h}a_h(\mathbf{k})
       + v_{h}b_h^\dagger(-\mathbf{k})
    \right)
,\quad\!\!
u_{h} \!=\! \biggl(\begin{array}{c}\!L_{h}\! \\ 
                              \!R_{h}\!
              \end{array}
        \biggr)\otimes \xi_{h}
\,,
\end{eqnarray}
where $\xi_{h}$ is the helicity two-eigenspinor,
$\hat h \xi_{h} = h \xi_{h}$. The Dirac equation then decomposes into
\begin{eqnarray}
{\rm i}\partial_{\eta}L_{h}-h|\mathbf{k}|L_{h} &=& m_{R}R_{h}+{\rm i}m_{I}R_{h}\,,
\label{LhRh}\\
{\rm i}\partial_{\eta}R_{h}+h|\mathbf{k}|R_{h} &=& m_{R}L_{h}-{\rm i}m_{I}L_{h}
\,.\nonumber
\end{eqnarray}

From $L_h$ and $R_h$, we can define the quantities
\begin{eqnarray}
 f_{0h} &=& |L_{h}|^2+|R_{h}|^{2},
\quad\;
 f_{3h} = |R_{h}|^2-|L_{h}|^2,
\label{f3asLR}\\
 f_{1h} &=& -2\Re(L_{h}R_{h}^{*}),
\qquad
 f_{2h} = 2\Im(L_{h}^{*}R_{h})
,
\nonumber
\end{eqnarray}
where $f_{0h}$ is the charge density, $f_{3h}$ the axial charge density,
$f_{1h}$ the scalar density and $f_{2h}$ the pseudoscalar density. Note that
one can easily show that $f_{0h}$ is conserved by Eq.~(\ref{LhRh}), reflecting
the charge conservation of the Dirac neutrinos.

The initial conditions corresponding to a particle number
$n_h(\mathbf{k})=|\beta_0|^2$ are
\begin{eqnarray}
\psi_{\mathbf{k}}=
  \left(\begin{array}{c}\alpha_0L_{h}^{+} + \beta_0L_{h}^{-}\\
                        \alpha_0R_{h}^{+} + \beta_0R_{h}^{-}
        \end{array}
  \right)
\,,\qquad
|\alpha_0|^{2} + |\beta_0|^{2}=1
\,,
\end{eqnarray}
where
\begin{eqnarray}
L_{h}^{+} &=& \sqrt{\frac{\omega(\mathbf{k})+h{k}}{2\omega(\mathbf{k})}}
\,,
\qquad\qquad
L_{h}^{-} = -i\frac{m}{|m|}
              \sqrt{\frac{\omega(\mathbf{k})-h{k}}{2\omega(\mathbf{k})}}\,,
\\
R_{h}^{+} &=& \frac{m^{*}}
                   {\sqrt{2\omega(\mathbf{k})(\omega(\mathbf{k})+h{k})}}
\,,
\qquad
R_{h}^{-} = i\frac{|m|}{\sqrt{2\omega(\mathbf{k})(\omega(\mathbf{k})-h{k})}}\,,
\nonumber
\end{eqnarray}
and $\omega(\mathbf{k}) =\sqrt{\mathbf{k}^2+|m|^2}$.
Since we assume to have initially zero neutrinos, we take $\beta_0=0$
in the following.

When $\phi_u^0$ ceases to oscillate, the particle number is given by
\begin{equation}
 n_h(\mathbf{k}) = \frac{1}{2\omega(\mathbf{k})}\left( h{k}f_{3h}
                                           + m_{R}f_{1h}
                                           + m_{I}f_{2h}
                                      \right)
                + \frac{1}{2}
\,.
\label{particle-number:fermions:kin}
\end{equation}
Of course there is no charge asymmetry, since there is an opposite amount of
antiparticles. However, when $m_I\not=0$, an asymmetry in the number of
particles with positive ($h=+$) and negative ($h=-$) helicity may be generated.
Note that in the limit $m_R,m_I\rightarrow 0$,
$n_{\mathbf{k}h}=\frac 12 hf_{3h}+ \frac 12$, since then chirality and
helicity coincide. Therefore,
\begin{equation}
2\left(n_+ - n_- \right)=f_{3+}+f_{3-}
\end{equation}
when the masses vanish. The factor \emph{two} on the left hand side occurs
because the total axial asymmetry gets contributions from particles and
antiparticles, while $n_h(\mathbf{k})$ counts just the particles.

\begin{figure}[htbp]
\epsfig{file=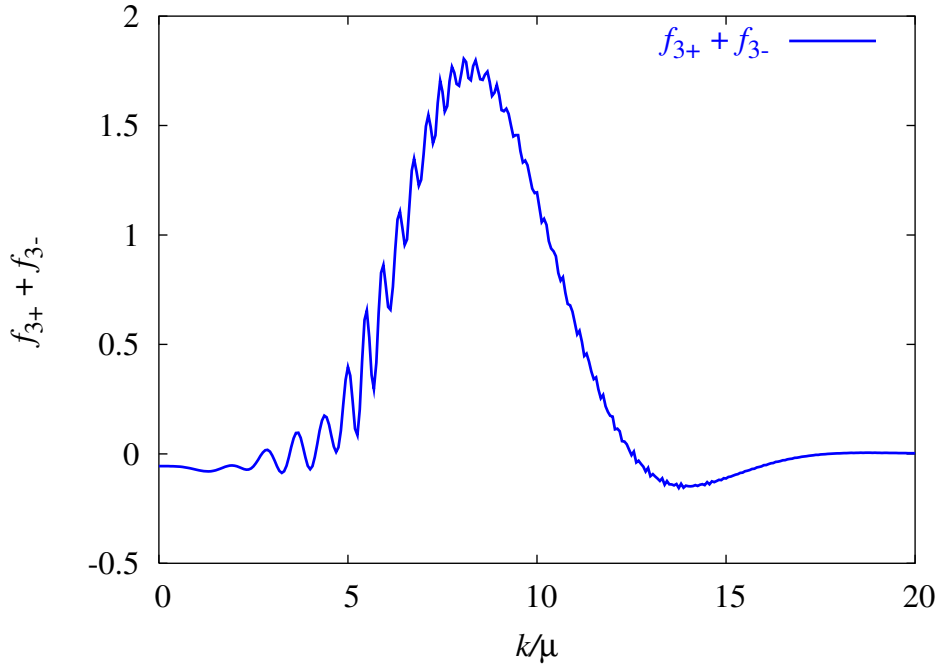,width=13cm}
\caption{\label{f3_k}
The axial asymmetry plotted over momentum $k$, which is taken to be
the physical momentum at the time when coherent oscillations begin.
The choice of parameters is: $\lambda_\nu=0.2$, $A^R=A^I=20\mu$,
$m=0.05\mu$, $\Gamma=0.002\mu$. 
}
\end{figure}

With a prime denoting a derivative \emph{w.r.t.} $\eta$, the scalar
equation of motion reads
\begin{equation}
\phi^{\prime\prime}+2\frac{a^\prime}{a}\phi^\prime
+a^2\frac{dV}{d\phi}+a \Gamma \phi^{\prime}=0\,.
\end{equation}
During radiation expansion, $a=a_R\eta$, and when
$H=a^\prime/a^2\ll \sqrt{\mu^2 \pm m^2}$, the solution to this equation is
well approximated by
\begin{equation}
\label{phiu0}
\phi_u^0\approx
\left[
A^R \cos\left(\sqrt{\mu^2-m^2}\frac{a_R}{2}\eta^2\right)+
{\rm i}
A^I \cos\left(\sqrt{\mu^2+m^2}\frac{a_R}{2}\eta^2\right)
\right]
(a_R\eta)^{-3/2}{\rm e}^{-\frac 14 \Gamma a_R \eta^2}\,,
\end{equation}
with the same assumptions for the real and imaginary parts as in the
previous section.
We use this solution to obtain the Dirac neutrino mass term~(\ref{m:Phi})
and numerically solve Eq.~(\ref{LhRh}) by integrating up to
the time when $\Gamma>H$, such that the Dirac mass term ceases to oscillate
and the axial charges
$f_{3h}(\mathbf{k})$ get frozen in. A typical plot of the spectrum of the
generated charge charge asymmetry is given in FIG.~\ref{f3_k}.
Particle production occurs at a time $t_{\rm Res}$ when the fermionic mode is
in resonance with the coherently oscillating field. The production of
the soft modes with small momentum $k$ is suppressed because the initial charge
asymmetry in the scalar field is small $(m^2/\mu) t_{\rm Res}\ll 1$, {\it cf.}
Eq.~(\ref{Asymmetry:Perturbative}). Consequently, the production of
asymmetry within modes which resonate later becomes stronger first. Eventually,
there is a damping effect due to the red-shifting of the oscillating
condensate and finally due to its decay at the rate $\Gamma$. Note that
due to Pauli blocking $-2\leq f_{3+}(\mathbf{k})+f_{3-}(\mathbf{k})\leq2$.

Of course, the axial asymmetry vanishes
in the case when the scalar charge~(\ref{charge:scalar}) is \emph{zero}.
When $A^I=0$, the term $\propto\gamma^5$ in the
Dirac equation~(\ref{Dirac:Eqn}) vanishes and there is obviously no
$CP$-violation. When $A^I\not=0$ but $m=0$ there is also \emph{zero}
scalar charge. Since then the phase is constant,
$\partial_\eta {\rm arg}(m_r+{\rm i}m_I)=0$, 
the $\gamma^5$-term can in principle be removed at all times
by a rephasing of the fermionic field. Consequently, even if we
do not perform this rephasing, we expect to find
$f_{3+}(\mathbf{k})+f_{3-}(\mathbf{k})\equiv 0$,
which can also be verified numerically.

The axial charge density stored within the neutrinos
is is the integral over the asymmetry within the modes
\begin{equation}
\label{Asymmetry:Resonant}
Q_A=\int\frac{d^3k}{(2\pi)^3}\left(
f_{3+}(\mathbf{k})+f_{3-}(\mathbf{k})
\right)\,,
\end{equation}
where $k$ is to be understood as the physical momentum at the time when
coherent oscillations begin.
The axial asymmetry to entropy ratio then turns out to be
\begin{equation}
\frac{n(\nu_L)-n(\bar\nu_L)}{s}=\alpha\frac{Q_A}{(\mu m_{Pl})^{3/2}}\,,
\end{equation}
which we want to compare with the perturbative
result~(\ref{Asymmetry:Perturbative}).

\begin{figure}[htbp]
\epsfig{file=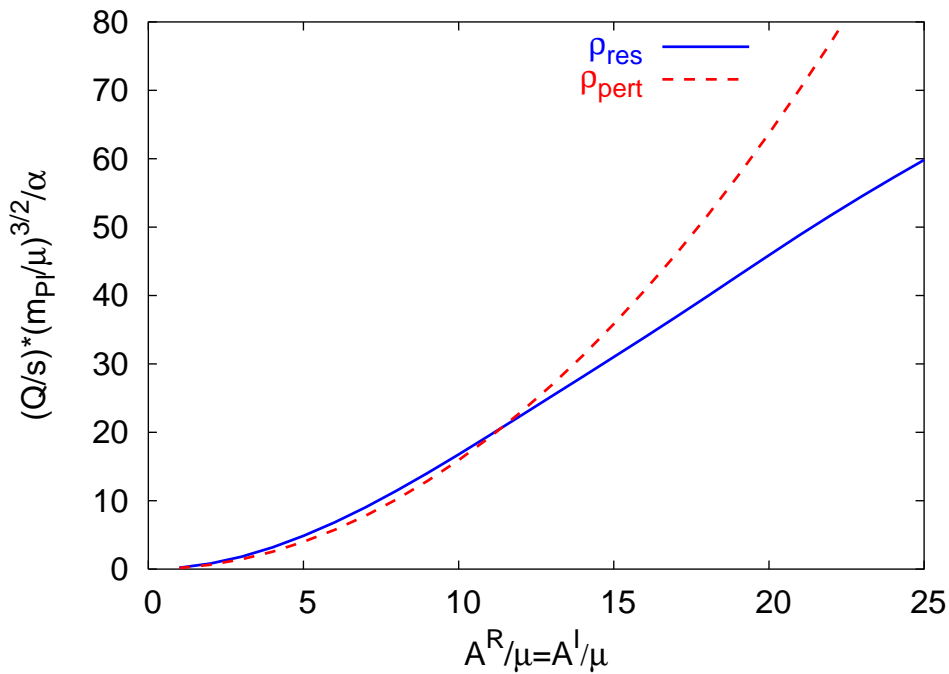,width=13cm}
\caption{\label{q_l:1}
The axial asymmetry plotted over the initial
amplitude of $\phi_u^0=A^R+{\rm i}A^I$.
The choice of parameters is $\lambda_\nu=0.2$,
$m=0.05\mu$, $\Gamma=0.01\mu$. 
}
\end{figure}

\begin{figure}[htbp]
\epsfig{file=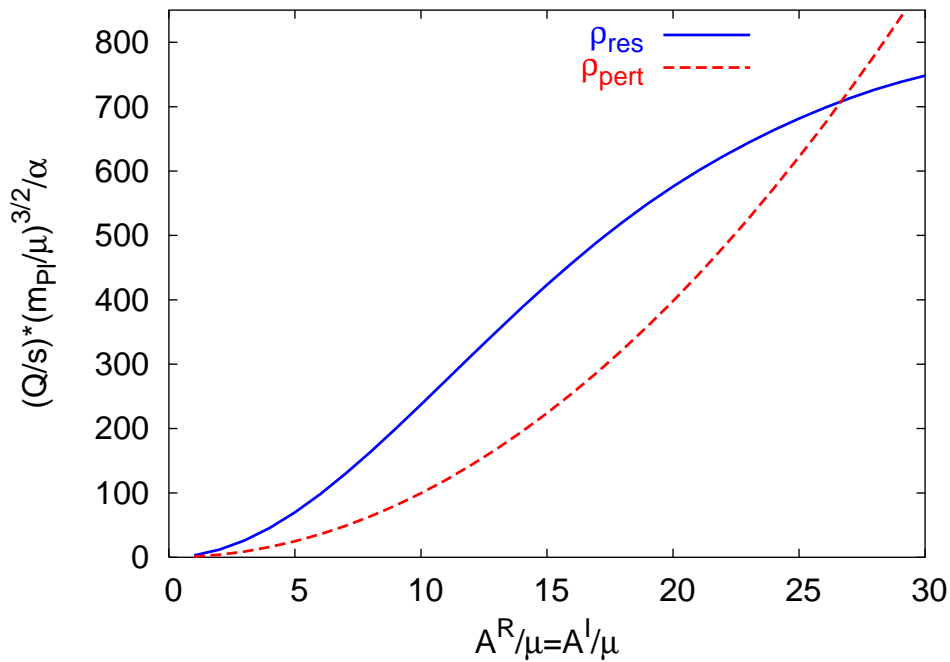,width=13cm}
\caption{\label{q_l:3}
The axial asymmetry plotted over the initial
amplitude of $\phi_u^0=A^R+{\rm i}A^I$.
The choice of parameters is $\lambda_\nu=0.2$,
$m=0.05\mu$, $\Gamma=\Gamma_\nu=0.0016\mu$.
}
\end{figure}

We denote the axial densities $n(\nu_L)-n(\bar\nu_L)$ by
$\rho_{\rm res}$ for the nonperturbative or resonant case of
Eq.~(\ref{Asymmetry:Resonant}) and
by $\rho_{\rm pert}$ for the perturbative decay as expressed in
Eq.~(\ref{Asymmetry:Perturbative}). The initial
amplitudes of the scalar field are chosen to be $A^R=A^I$.
We display the produced
axial asymmetries over the initial amplitudes
in FIGs.~\ref{q_l:1} and~\ref{q_l:3}, where
we have taken different values for the damping rate $\Gamma$.

Clearly, the perturbative source $\rho_{\rm pert}$ gets enhanced by the
factor $\Gamma_\nu/\Gamma$ in~(\ref{Asymmetry:Perturbative})
as the damping $\Gamma$ becomes smaller.
Note that we have consistently chosen $\Gamma_\nu \leq \Gamma$ with
the case $\Gamma_\nu=\Gamma$ displayed in FIG.~\ref{q_l:3}.
But also the nonperturbative contribution grows for smaller decay rates,
because coherent oscillations last longer and a larger phase space volume may
be filled as the fermionic modes are red-shifted. However,
while initially $\rho_{\rm pert}$ and $\rho_{\rm res}$
grow as the square of the initial scalar amplitude, $\rho_{\rm res}$ gets
suppressed
for large amplitudes due to Pauli blocking, which we do not take into account
in our formula for the perturbative asymmetry~(\ref{Asymmetry:Perturbative}).
In either case as displayed in FIGs.~\ref{q_l:1} and~\ref{q_l:3},
we note that the total asymmetry is the sum
of the individual contributions, $\rho_{\rm pert}+\rho_{\rm res}$. 

\section{Conclusions}
Nonperturbatively sourced single field baryogenesis
is a viable scenario. We have shown that it contributes over a wide range
of parameter space by the same order of magnitude as the perturbative source
to the baryon asymmetry. Note that a rescaling of the Yukawa coupling
$\lambda_\nu$ can be absorbed into different intitial amplitudes $A^{R,I}$,
such that the effect can be read of from FIGs.~\ref{q_l:1} and~\ref{q_l:3}.
Besides the model presented
here, the coherent baryogenesis mechanism is an example for generating the
bayon asymmetry directly from preheating.
We conclude that
processes of nonperturbative particle production may be of importance
for explaining the baryon asymmetry of the Universe.

\begin{acknowledgments}
We thank R.~H.~Brandenberger
and K.~Kainulainen for participation in early stages of this
work and A.~Notari for useful discussions.
The work of KB is funded by NSERC (Canada) and by the Fonds de Recherche
sur la Nature et les Technologies du Qu\'ebec.
\end{acknowledgments}

\end{document}